\documentclass{ws-procs9x6}

\usepackage{epic,eepic}
\usepackage{epsfig,cite,indentfirst}
\usepackage{multicol,boxedminipage}
\usepackage{varioref} 
\usepackage{amsmath}

\begin{document}

\title{XXZ scalar products, Miwa variables and discrete KP}

\author{O Foda and G Schrader}

\address{Department of Mathematics and Statistics, \\
         University of Melbourne, \\ 
         Parkville, Victoria 3010, Australia.\\
         E-mail: o.foda, g.schrader@ms.unimelb.edu.au}

\newcommand{\field}[1]{\mathbb{#1}}
\newcommand{\CC}{\field{C}}
\newcommand{\NN}{\field{N}}
\newcommand{\ZZ}{\field{Z}}

\newcommand{\B}{{\mathcal B}}
\newcommand{\D}{{\mathcal D}}
\newcommand{\M}{{\mathcal M}}
\newcommand{\R}{{\mathcal R}}
\newcommand{\T}{{\mathcal T}}
\newcommand{\U}{{\mathcal U}}
\newcommand{\Y}{{\mathcal Y}}
\newcommand{\Z}{{\mathcal Z}}

\newcommand{\tT}{\widetilde{\mathcal T}}
\newcommand{\tU}{\widetilde{\mathcal U}}

\renewcommand{\P}{{\mathcal P}}
\newtheorem{ca}{Figure}

\maketitle

\abstracts{
We revisit the quantum/classical integrable model correspondence
in the context of inhomogeneous finite length XXZ 
spin-$\frac{1}{2}$ chains with periodic boundary conditions and 
show that the Bethe scalar product of an arbitrary state and a Bethe 
eigenstate is a discrete KP $\tau$-function. The continuous Miwa 
variables of discrete KP are the rapidities of the arbitrary state.}

\def\ll{\left\lgroup}
\def\rr{\right\rgroup}

\newcommand{\Proof}{\medskip\noindent {\it Proof: }}
\def\proofend{\ensuremath{\square}}

\def\no{\nonumber}

\def\union{\mathop{\bigcup}}
\def\vac{|\mbox{vac}\rangle}
\def\cav{\langle\mbox{vac}|}

\def\lprod{\mathop{\prod{\mkern-29.5mu}{\mathbf\longleftarrow}}}
\def\rprod{\mathop{\prod{\mkern-28.0mu}{\mathbf\longrightarrow}}}

\def\r{\rangle}
\def\l{\langle}

\def\a{\alpha}
\def\b{\beta}
\def\g{\gamma}
\def\d{\delta}
\def\e{\epsilon}
\def\l{\lambda}

\def\s{\sigma}

\def\eps{\varepsilon}
\def\hb{\hat\beta}

\def\tg{\operatorname{tg}}
\def\ctg{\operatorname{ctg}}
\def\sh{\operatorname{sh}}
\def\ch{\operatorname{ch}}
\def\cth{\operatorname{cth}}
\def\th{\operatorname{th}}

\def\tla{\tilde{\lambda}}
\def\tmu{\tilde{\mu}}

\def\sul{\sum\limits}
\def\pl{\prod\limits}

\def\pd #1{\frac{\partial}{\partial #1}}
\def\const{{\rm const}}
\def\argum{\{\mu_j\},\{\la_k\}} 
\def\umarg{\{\la_k\},\{\mu_j\}} 

\def\prodmu #1{\prod\limits_{j #1 k} \sinh(\mu_k-\mu_j)}
\def\prodla #1{\prod\limits_{j #1 k} \sinh(\lambda_k-\lambda_j)}

\newcommand{\bl}[1]{\makebox[#1em]{}}

\def\tr{\operatorname{tr}}
\def\Res{\operatorname{Res}}
\def\det{\operatorname{det}}

\newcommand{\boldN}{\boldsymbol{N}}
\newcommand{\bra}[1]{\langle\,#1\,|}
\newcommand{\ket}[1]{|\,#1\,\rangle}
\newcommand{\bracket}[1]{\langle\,#1\,\rangle}
\newcommand{\infinity}{\infty}

\renewcommand{\labelenumi}{\S\theenumi.}

\let\up=\uparrow
\let\down=\downarrow
\let\tend=\rightarrow
\hyphenation{boson-ic
             ferm-ion-ic
             para-ferm-ion-ic
             two-dim-ension-al
             two-dim-ension-al
             rep-resent-ative
             par-tition}

\renewcommand{\mod}{\textup{mod}\,}
\newcommand{\wt}{\text{wt}\,}

\hyphenation{And-rews
             Gor-don
             boson-ic
             ferm-ion-ic
             para-ferm-ion-ic
             two-dim-ension-al
             two-dim-ension-al}

\setcounter{section}{0}

\section{Introduction.}

Quantum models of the statistical mechanical type (the only 
quantum models discussed in this note) such as integrable 
1-dimensional quantum spin chains, and classical models 
such as integrable nonlinear partial differential equations, 
are related 
in the sense that the methods used to compute in the former, 
particularly the quantum inverse scattering transform, also 
known as the algebraic Bethe Ansatz, are quantum versions 
of those used to compute in the latter, namely the classical 
inverse scattering transform. It is therefore natural to 
expect that the quantum integrable models have classical 
limits in which they reduce to classical counterparts.

What is less than natural to expect, at least to our minds, 
is that basic objects in quantum integrable models, such as 
the correlation functions, turn out to have direct interpretations 
in terms of objects in classical integrable models, such as 
solutions of integrable nonlinear partial differential and 
difference equations, {\it without taking a classical limit}. 
But this turns out to be the case, and it points to a direct 
connection between quantum and classical integrable models 
that is distinct from, and to our minds at least as fundamental 
as that obtained by taking a classical limit.  

\paragraph{Notes on the literature.} The following is far from 
a comprehensive survey of the relevant literature. To the best 
of our knowledge, a direct connection between quantum (statistical 
mechanical) and classical models of the type that we are interested 
in first appeared in \cite{wu-mccoy}, where Ising spin-spin 
correlation functions in the scaling limit were shown to satisfy 
Painlev\'{e} equation of the third kind, and subsequently in 
\cite{perk, au-yang-perk}, where critical Ising correlation 
functions on the lattice were shown to satisfy the Toda lattice 
equation in Hirota's bilinear form. Further results, along 
the same lines as in \cite{wu-mccoy}, for the XXZ spin chain 
at the free fermion point, were obtained in \cite{korepin}, 
as reviewed in \cite{korepin-book}.

The fact that $\tau$-functions (solutions of Hirota's bilinear 
equations) appear in the Ising model as well as in KP theory 
was discussed in works by the Kyoto group and reviewed in 
\cite{miwa-review} where it was argued that the mathematical 
reason underlying this coincidence is the fact that both 
quantum and classical models are based on infinite dimensional 
Lie algebras that are realized in terms of free fermions. 

Closest to the spirit of this note is the work of Krichever 
{\it et al.} \cite{krichever}, reviewed in \cite{zabrodin}. 
The starting point of \cite{krichever} is the observation 
that the Bethe eigenvalues satisfy Hirota's difference 
equation, various limits of which lead to a large number 
of integrable differential and difference equations 
\cite{hirota}. We will comment on the results of 
\cite{krichever} and how they differ from the result in this 
note in section {\bf 6}. More recently, studies of the 
ultra-discrete limit of quantum integrable spin chains 
revealed many classical integrable structures \cite{kuniba}.

Finally, while we are only interested in integrable quantum 
models in statistical mechanics in this note, it is important 
to mention bosonisation (the operator formulation of Sato's 
theory) as a deep and established correspondence between 
the quantum field theories of free fermions, which are 
integrable quantum models, and classical integrable 
hierarchies, as reviewed in \cite{djkm}. 
In this correspondence, expectation values of fermion operators 
have direct interpretations in terms of solutions of integrable 
nonlinear partial differential equations. Bosonization was 
further extended to connect KP theory and conformal field 
theories on Riemann surfaces (which are integrable quantum 
models) in \cite{kawamoto}.

The long term aim of 
our work is to develop a correspondence between integrable 
statistical mechanical models and classical integrable 
hierarchies that is as direct and detailed as that obtained 
by bosonisation between free fermions and classical integrable 
hierarchies.

\paragraph{Bethe scalar products and continuous KP $\tau$-functions.}
Consider the inhomogeneous length-$L$ XXZ spin-$\frac{1}{2}$ chain 
with periodic boundary conditions. Following \cite{fwz}, the Bethe 
scalar product 
$\langle 
\lambda_1, \cdots, \lambda_N| \mu_1, \cdots, \mu_N 
\rangle_{\beta}$ 
of an arbitrary state 
$\langle \lambda_1, \cdots, \lambda_N|$ 
where the auxiliary space rapidities 
$\{\lambda_1, \cdots, \lambda_N\}$ 
are free, and a Bethe eigenstate 
$|\mu_1, \cdots, \mu_N \rangle_{\beta}$ 
where the auxiliary space rapidities 
$\{\mu_1, \cdots,\mu_N\}$ 
obey the Bethe equations, is a polynomial 
$\tau$-function of the continuous (differential) KP hierarchy. 
In this identification, the KP time 
variables $\{t_1, t_2, \cdots\}$ are power sums of the free 
rapidities $\{\lambda_1, \cdots, \lambda_N\}$. However, these 
polynomial KP $\tau$-functions involve by construction more 
time variables than free rapidities. The reason is as follows. 

Expanding the scalar product in terms of Schur polynomials 
$s_{\lambda}$, associated to Young diagrams $\{\lambda\}$, 
that are functions of the rapidities 
$\{\lambda_1, \cdots, \lambda_N\}$,
the maximal number of rows in any Young diagram $\lambda$ 
is $N$. 
Switching to KP time variables $\{t_1, t_2, \cdots\}$ that 
are powers sums in the rapidities, we obtain character polynomials 
$\chi_{\lambda}$ that depend on effectively as many time variables 
as the number of cells in (that is, the size of) $\lambda$ which 
is larger than $N$. Consequently, 
the KP time variables $\{t_1, t_2, \cdots\}$ were {\it formally}  
considered in \cite{fwz} to be independent, and the Bethe scalar 
product was 
defined as a {\it restricted} KP $\tau$-function obtained by 
setting $\{t_1, t_2, \cdots\}$ to be power sums of a smaller 
number of independent variables $\{\lambda_1, \cdots, \lambda_N\}$.

\paragraph{Bethe scalar products and discrete KP $\tau$-functions.}
In this note, we simplify the correspondence of \cite{fwz} by working 
solely in terms of the free rapidities $\{\lambda_1, \cdots, \lambda_N\}$ 
which are now continuous Miwa variables and the $\tau$-functions that 
we obtain are those of the discrete KP hierarchy \cite{djm2, ohta}.

\paragraph{Outline of contents.} 
In section {\bf 2}, we recall basic facts related to symmetric functions, 
Casoratian matrices and Casoratian determinants.
In {\bf 3}, we recall basic facts related to the XXZ spin-$\frac{1}{2}$ 
chain, the algebraic Bethe Ansatz,
the Bethe scalar product, recall Slavnov's determinant 
expression of the Bethe scalar product and show that it is 
a Casoratian determinant.
In {\bf 4}, we recall basic facts related to the continuous 
and discrete KP hierarchies and define the Miwa variables that 
relate the two.
In {\bf 5}, we show that Bethe scalar products in the XXZ 
spin-$\frac{1}{2}$ chain with periodic boundary conditions 
are discrete KP $\tau$-functions.
In {\bf 6}, we collect a number of remarks.
Space limitations allow us to give no more than 
the minimal definitions necessary to fix the notation and terminology
supplemented by references to relevant sources.

\section{Symmetric functions and Casoratians.}
\label{functions}

\noindent The canonical reference to symmetric functions is 
\cite{macdonald-book}. 
Casoratian matrices and determinants are carefully discussed 
in \cite{ohta}. The definitions in \cite{ohta} are more 
general than those used in this note.

\paragraph{Frequently used notation.} We use $\{x\}$ 
for the set of finitely many variables $\{x_1, x_2, \cdots, x_N\}$, 
or infinitely many variables 
$\{x_1, x_2, \cdots \}$. The cardinality of the set should
be clear from the context. We use 
$\{\widehat{x}_m\}$ for $\{x\}$ but with 
the element $x_m$ missing. In the case of sets with 
a repeated variable $x_i$, we use the superscript 
$(m_i)$ to indicate the multiplicity of $x_i$, as 
in $x_i^{(m_i)}$. For example,
$\{x_1^{(3)}, x_2, x_3^{(2)}, x_4, \cdots \}$ 
is the same as
$\{x_1, x_1, x_1, x_2, x_3, x_3, x_4, \cdots \}$
and $f\{\cdots, x_{i}^{(m_i)}, \cdots\}$ 
is equivalent to saying that $f$ depends on $m_i$ distinct 
variables all of which have the same value $x_i$. 
For simplicity, we use $x_i$ to indicate $x_i^{(1)}$. 
In calculations, it is safer to think of any $x_i$ with 
multiplicity $m_i > 1$ initially as distinct, that is
$\{x_{i, 1}, x_{i, 2}, \cdots, x_{i, m_i}\}$, then set
these $m_i$ variables equal to the same value $x_i$ at 
the end.

We use the bracket notation 
$[x] = e^x - e^{-x}$, and 
\begin{equation} 
\Delta\{x\}              = 
\prod_{1 \leq i < j \leq N} \left( x_i       - x_j \right), 
\quad
\Delta_{trig}\{\lambda\} = 
\prod_{1 \leq i < j \leq N} \left[ \lambda_i - \lambda_j \right]
\label{vandermonde}
\end{equation}
\noindent for the Vandermonde determinant and its trigonometric analogue.

\paragraph{The elementary symmetric function $e_i\{x\}$} in $N$ 
variables $\{x\}$ is the coefficient of $k^i$ in the expansion 

\begin{equation}
\prod_{i=1}^{N}
\left( 1 + x_i \, k \right)
=
\sum_{i=0}^{\infty}
e_i\{x\}
\, 
k^i
\label{elementary-symmetric}
\end{equation}

\noindent For example, $e_0\{x\} = 1$, $e_1(x_1,x_2,x_3) = x_1+x_2+x_3$,
$e_2(x_1,x_2) = x_1x_2$. $e_i\{x\} = 0$, for $i<0$ and for 
$i>N$.

\paragraph{The complete symmetric function $h_i\{x\}$} in $N$ 
variables $\{x\}$ is the coefficient of $k^i$ in the expansion 

\begin{equation}
\prod_{i=1}^{N}
\frac{1}{1-x_i \, k}
=
\sum_{i=0}^{\infty}
h_i\{x\}
\, 
k^i
\label{complete-symmetric}
\end{equation}

\noindent For example, $h_0\{x\} = 1$, $h_1(x_1,x_2,x_3) = x_1+x_2+x_3$,
$h_2(x_1,x_2) = x_1^2+x_1x_2+x_2^2$, and 
$h_{i}\{x\} = 0$ for $i<0$. 

\paragraph{Useful identities for $h_i\{x\}$.} From Equation 
(\ref{complete-symmetric}), it is straightforward to show that

\begin{equation}
h_i\{x\} = h_i\{\widehat{x}_m\} + x_m h_{i-1}\{x\}
\label{i1}
\end{equation}

\noindent and from that, one obtains

\begin{multline}
h_i\{x_1, x_2, \cdots, x_N\} = 
\\
h_i \{x_1^{(2)}, x_2, \cdots, x_N\} 
- x_1 
h_{i-1} \{x_1^{(2)}, x_2, \cdots, x_N\} 
\label{i2}
\end{multline}

\begin{multline}
(x_2 - x_1) h_i\{x_1^{(2)}, x_2^{(2)}, x_3, \cdots, x_N\} = \\
x_2 h_i\{x_1,       x_2^{(2)}, \cdots, x_N\} -
x_1 h_i\{x_1^{(2)}, x_2,       \cdots, x_N\} 
\label{i3}
\end{multline}

\paragraph{The discrete derivative $\Delta_m h_i\{x\}$} of $h_i\{x\}$ 
with respect to any one variable $x_m \in \{x\}$ is defined using 
Equation (\ref{i1}) as  

\begin{equation} 
\Delta_{m} h_i\{x\} = 
\frac{
h_i\{x\} - h_i\{ \widehat{x}_m \}
}
{x_m
} 
= h_{i-1}\{x\}
\label{discrete-derivative}
\end{equation}

\noindent Note that the effect of applying $\Delta_m$ to
$h_i\{x\}$ is a complete symmetric function $h_{i-1}\{x\}$ 
of degree $i-1$ in the same set of variables $\{x\}$.
The difference operator in Equation (\ref{discrete-derivative})
is not the most general definition of a discrete derivative, 
but it is sufficient for the purposes of this note. For a more
general definition, see \cite{ohta}.

\paragraph{The Schur polynomial $s_{\lambda}\{x\}$} indexed by 
a Young diagram $\lambda=[\lambda_1,\ldots,\lambda_r]$ with 
$\lambda_{i} \neq 0$, for $1   \leq i \leq r$,
and 
$\lambda_{i} =    0$, for $r+1 \leq i \leq N$, 
is

\begin{equation}
s_{\lambda}\{x\}
=
\frac{
\det
\ll
x_i^{\lambda_j-j+N}
\rr_{1 \leq i,j \leq N}
}
{
\Delta\{x\}
}
=
\det
\ll
h_{\lambda_i-i+j}\{x\}
\rr_{1\leq i,j \leq N}
\label{Schur}
\end{equation}

\noindent For example, $s_{\phi} \{x\} =1$,
$s_{[1  ]}(x_1,x_2,x_3) = x_1+x_2+x_3$,
$s_{[1,1]}(x_1,x_2    ) = x_1 x_2$.
The first equality in Equation (\ref{Schur}) is the definition of 
$s_{\lambda}\{x\}$. 
The second is the {\it Jacobi-Trudi identity} for $s_{\lambda}\{x\}$.
$s_{\lambda}\{x\}$ is symmetric in the elements of $\{x\}$ 
and requires no more than $r$ (the number of non-zero rows in $\lambda$) 
variables to be non-vanishing.

\paragraph{The one-row character polynomial $\chi_{i}\{t\}$} indexed 
by a one-row Young diagram of length $i$, is the $i$-th coefficient 
in the generating series

\begin{equation}
\sum_{i=0}^{\infty}
\chi_{i}\{t\}
\, 
k^i
=
\exp
\ll
\sum_{i=1}^{\infty}
t_i \, k^i
\rr
\label{one-row-character}
\end{equation}

\noindent For example,
$\chi_0\{t\} = 1$,
$\chi_1\{t\} = t_1$,
$\chi_2\{t\} = \frac{t_1^2}{2}+t_2$,
$\chi_3\{t\} = \frac{t_1^3}{6}+ t_1 t_2 + t_3$,
and $\chi_{i}\{t\} = 0$ for $i<0$. Since $t_i$ has degree $i$, 
$\chi_i$ is not symmetric in $\{t\}$ and generally depends on 
as many $t$-variables as the row-length $i$. 

\paragraph{The character polynomial $\chi_{\lambda}\{t\}$} 
indexed by a Young diagram 
$\lambda = [\lambda_1, \lambda_2, \ldots, \lambda_r]$
with $r$ non-zero-length rows, $r \leq N$, is

\begin{equation}
\chi_{\lambda}\{t\}
=
\det
\ll
\chi_{\lambda_i-i+j}\{t\}
\rr_{1\leq i,j \leq n}
\label{r-row-character}
\end{equation}

\noindent For example
$\chi_{[1,1]}\{t\} = \frac{t_1^2}{ 2}-t_2$,
$\chi_{[2,1]}\{t\} = \frac{t_1^3}{ 3}-t_3$,
$\chi_{[2,2]}\{t\} = \frac{t_1^4}{12}-t_1 t_3 + t_2^2$.
Notice that $\chi_{\lambda}\{t\}$ can depend on all $t_i$, 
for $i \leq |\lambda|$, where $|\lambda|$ is the sum of 
the lengths of all rows in (or area of) $\lambda$. 

\paragraph{From character polynomials to Schur polynomials.}
Assuming that the $t$-variables are independent and that we
have sufficiently many $x$-variables, then setting 
$ t_m        \rightarrow \frac{1}{m} \sum_{i=1}^{N} x_i^m$
sends
$\chi_i\{t\} \rightarrow                h_i \{x\}$.
In this note, as in \cite{fwz}, we study Bethe scalar products 
that are polynomials in $N$ variables $\{x_1, x_2, \cdots, x_N\}$. 
We can expand these scalar products in terms of 
Schur polynomials $s_{\lambda} \{x\}$ where 
$\{\lambda\}$ has at most $N$ rows,
or in terms of the corresponding character polynomials 
$\chi_{\lambda} \{t\}$
that require more $t$-variables (which are power sums in 
the $x$-variables) than $N$ and 
therefore cannot be all independent.
We choose to work in terms of the $x$-variables and
$s_{\lambda} \{x\}$.

\paragraph{Casoratian matrices and determinants.} 
A Casoratian matrix $M$ of the type that appears in this note
is such that the elements $M_{ij}$ satisfy either 

\begin{equation}
M_{i,   j+1}\{x\} = \Delta_m M_{ij}\{x\}, \quad \text{or} \quad 
M_{i+1, j  }\{x\} = \Delta_m M_{ij}\{x\}
\label{casoratian-condition}
\end{equation}

\noindent where the discrete derivative $\Delta_m$ is taken with 
respect to any one variable $x_m \in \{x\}$. 
If $M$ is a Casoratian matrix, then $\det M$ is a Casoratian 
determinant. Casoratian determinants are discrete analogues 
of Wronskian determinants.

\section{The XXZ spin-$\frac{1}{2}$ chain and the Algebraic Bethe Ansatz.}

\noindent The XXZ spin-$\frac{1}{2}$ chain is discussed in detail 
in \cite{baxter-book, jimbo-miwa-book}. A standard reference to 
the algebraic Bethe Ansatz, including the Bethe scalar product 
and Slavnov's determinant expression, is \cite{korepin-book}. 
We leave the definition of auxiliary and quantum spaces, auxiliary 
and quantum rapidities, and the precise action of the various 
operators to \cite{korepin-book}. 

\paragraph{Frequently used variables.} In the following, $L$ is 
the number of sites in a periodic XXZ spin-$\frac{1}{2}$ chain, 
and $N$ is the number of Bethe operators $B(\mu_i)$ that act on 
the reference state $|0\rangle$ to create an XXZ state 
$| \mu_1, \cdots, \mu_N \rangle$.
$N$ is also the rank of the matrix whose determinant is Slavnov's 
expression for the Bethe scalar product.
We use the set $\{\lambda\}$ for the free auxiliary 
space rapidities,
$\{\mu\}$ or more explicitly $\{\mu_{\beta}\}$ for the auxiliary 
space rapidities that satisfy the Bethe equations, and $\{\nu\}$ 
for the quantum space rapidities (the inhomogeneities). A Bethe 
eigenstate state whose rapidities satisfy the Bethe equations 
is also denoted by a subscript $\beta$, such as $| \lambda \rangle_{\beta}$.
$\gamma$ is the crossing parameter.
We use the exponentiated variables 
$\{x_i, y_i, z_i, q\}$ 
$=$
$\{e^{\lambda_i}, e^{\mu_i}, e^{\nu_i}, e^{\gamma}\}$, 
but still refer to the exponentiated variables $\{x, y, z\}$ 
as rapidities rather than exponentiated rapidities 
for simplicity.

\paragraph{The $L$-operator} of the XXZ spin-$\frac{1}{2}$ chain 
is 

\begin{equation}
L_{ai}(\lambda, \nu)
=
\ll
\begin{array}{cccc}
[\lambda - \nu + \gamma] & 0 & 0  & 0 \\
 0 & [\lambda - \nu] & [\gamma]   & 0 \\
 0 & [\gamma] & [\lambda - \nu]   & 0 \\
 0 & 0 & 0 & [\lambda - \nu + \gamma]
\end{array}
\rr_{ai}
\label{L-operator}
\end{equation}

\noindent where $a$ is an auxiliary space index 
and $i$ is a quantum space index.

\paragraph{The monodromy matrix} of the inhomogeneous length-$L$ 
XXZ spin-$\frac{1}{2}$ chain is 

\begin{equation}
T_a(\lambda)
=
\ll
\begin{array}{cc}
A(\lambda) & B(\lambda)
\\
C(\lambda) & D(\lambda)
\end{array}
\rr_{a}
=
\prod_{i=1}^{L} L_{a i}(\lambda, \nu_i) 
\label{monodromy}
\end{equation}

\noindent where it is conventional to suppress the dependence on 
the inhomogeneous quantum space rapidities 
$\nu_i$ in $T_a$ and its elements, and each of the operators $A$, 
$B$, $C$, and $D$ acts in the tensor product 
$V_1 \otimes \cdots \otimes V_L$ where $V_i$ is a vector space 
isomorphic to $\mathbb{C}^2$. 

\paragraph{The transfer matrix} is the trace of the monodromy 
matrix over the auxiliary space, 

\begin{equation}
{\rm Tr}_a T_a(\lambda) = A(\lambda) + D(\lambda)
\label{transfer-matrix}
\end{equation}

\paragraph{An arbitrary state $| \mu \rangle $} is 
generated by the action of the $B(\mu)$ operators on the 
reference state 
$|0 \rangle = \otimes^{L} \ll {1\atop 0}\rr$,

\begin{equation}
|\mu \rangle = B(\mu_1) \ldots B(\mu_N) |0 \rangle
\label{out-state}
\end{equation}

\paragraph{An arbitrary dual state $ \langle \lambda |$} 
is generated by the action of the $C(\lambda)$ operators on 
the dual reference state $\langle 0| = \otimes^{L} \ll 1\ 0\rr$, 

\begin{equation}
\langle \lambda| = \langle 0| \, C(\lambda_1) \ldots C(\lambda_N)
\label{in-state}
\end{equation}

\paragraph{The scalar product} of a state and a dual state is

\begin{equation}
\langle \lambda | \mu \rangle
=
\langle 0|
\, C(\lambda_1) \ldots C(\lambda_N)\, B(\mu_1)\ldots B(\mu_N)|0
\rangle
\label{scalar-product}
\end{equation}

\paragraph{A Bethe eigenstate $| \mu \rangle_{\beta}$} is 
an eigenstate of the transfer matrix,

\begin{equation}
\ll
A(\lambda) + D(\lambda)
\rr
| \mu \rangle_{\beta} =
E(\lambda) |\mu \rangle_{\beta}
\label{bethe-eigenstate}
\end{equation}

\noindent where $E(\lambda)$ is the corresponding Bethe eigenvalue. 
For a state $|\mu\rangle$ to be a Bethe eigenstate, its auxiliary 
space rapidities must satisfy a set of Bethe equations.

\paragraph{The Bethe equations} that must be satisfied by the 
$N$ auxiliary space rapidities of a state 
$| \mu \rangle = B(\mu_1) \cdots B(\mu_N) | 0 \rangle$
in order to be a Bethe eigenstate, in the specific case of the 
inhomogeneous length-$L$ spin-$\frac{1}{2}$ chain, are

\begin{equation}
\frac{\prod_{i=1}^{L} [\mu-\nu_i+\gamma]}{\prod_{i=1}^{L} [\mu-\nu_i]}
\prod_{j \not= i}^{N}
\frac{[\mu_i - \mu_j - \gamma]}{[\mu_i - \mu_j + \gamma]}
=
1
\label{bethe}
\end{equation}

\noindent where $\{\nu_1, \cdots, \nu_L\}$, are the quantum space
rapidities, which are taken to be part of the parameters that specify 
the spin chain, rather than the definition of the Bethe state.

\paragraph{A Bethe scalar product} is a scalar product of an arbitrary
state $\langle \lambda |$ and a Bethe eigenstate $|\mu \rangle_{\beta}$, 

\begin{equation}
\langle \lambda | \mu \rangle_{\beta}
=
\langle 0|
\, 
C(\lambda_1)      \ldots C(\lambda_N)
\, 
B(\mu_{1, \beta}) \ldots B(\mu_{N, \beta})|0 \rangle
\label{bethe-scalar-product}
\end{equation}

\noindent Bethe scalar products as in Equation (\ref{bethe-scalar-product}) 
play a central role in computing XXZ correlation functions \cite{kitanine}, 
hence their importance. 

\paragraph{Slavnov's determinant expression.} In \cite{slavnov}, Slavnov 
obtained an elegant determinant expression for the Bethe scalar product, 

\begin{equation}
\langle \lambda | \mu \rangle_{\beta}
=
[\gamma]^N
\frac{
\prod_{i,j=1}^{N}[ \lambda_i - \mu_j + \gamma]
}
{
\Delta\{\lambda\} 
\Delta\{\mu    \} 
}
\prod_{k=1}^{N}\prod_{l=1}^{L}
\, [\lambda_k-\nu_l] \, [\mu_k-\nu_l]
\, \det
\Omega
\label{slavnov1}
\end{equation}

\noindent where the components of the $N$$\times$$N$ matrix $\Omega$ 
are 

\begin{multline}
\Omega_{ij} = \frac{1}{[\lambda_i-\mu_j] [\lambda_i-\mu_j+\gamma]} 
\ 
-
\\
\frac{1}{[\mu_j-\lambda_i] [\mu_j-\lambda_i+\gamma]}
\prod_{k=1}^{L}
\frac{[\lambda_i-\nu_k+\gamma]}{[\lambda_i-\nu_k]}
\prod_{l=1}^{N}
\frac{[\lambda_i - \mu_l - \gamma]}
     {[\lambda_i - \mu_l + \gamma]}
%\label{components}
\end{multline}

\noindent Slavnov's scalar product is the main object of interest 
in this note. We wish to show that it is a Casoratian determinant 
and that the latter satisfy the bilinear identities of a discrete 
KP hierarchy \cite{ohta}. 

\paragraph{Re-writing Slavnov's determinant expression.} In \cite{fwz}, 
it was found useful to rewrite Slavnov's determinant expression for 
the Bethe scalar product as follows. First, we change variables and
work in terms of exponentials of the original variables as follows

\begin{equation}
\label{set}
\{
e^{2\lambda_i},
e^{2    \mu_i},
e^{2    \nu_i},
e^{    \gamma}
\}
\rightarrow
\{
x_i, y_i, z_i, q
\}
\end{equation}

\noindent but continue to call the exponentials $\{x, y, z\}$ rapidities 
as that is simpler and should cause no confusion. Ignoring prefactors that 
do not depend on $\{x\}$, it was shown in \cite{fwz} that the relevant part 
of Slavnov's determinant expression can be re-written as 

\begin{equation}
\det \Omega' = \frac{\det \Omega}{\Delta\{x\}},
\ \
\textrm{where}
\ \
\mathbf{\Omega}_{ij}
=
\sum_{k=1}^{N+L-1} 
x_i^{k-1} \, \kappa_{kj},
\quad
\kappa_{kj} = -\sum_{l=1}^{k} y_j^{l-k-1} \, \rho_{lj}, 
\label{needed}
\end{equation}

\noindent and

\begin{eqnarray}
\rho_{lj} 
&=&
\ll
\prod_{m=1}^{L}
\left(y_j q -z_m q^{-1} \right)
\rr
\ll
\prod_{n\not=j}^{N}
\left(y_j - y_n q^2 \right)
\cdot
e_{(L+N-l)} \{-\widehat{y}_j q^{-2} \}\{-z \}
\rr
\nonumber
\\
&-&
\ll
\prod_{m=1}^{L}
\left(y_j q -z_m q \right)
\rr
\ll
\prod_{n\not=j}^{N}
\left(y_j  - y_n q^{-2}\right)
\cdot
e_{(L+N-l)} \{ -\widehat{y}_j q^2 \}\{-z q^{-2} \}
\rr
\nonumber
\\
\label{rho}
\end{eqnarray}
\noindent In Equation (\ref{rho}), $e_k \{\widehat{y}_j\} \{z\}$ is 
the $k$-th elementary symmetric polynomial in the set of variables 
$\{y\}\cup \{z\}$ with the omission of $y_j$.  

\smallskip
\paragraph{A Bethe scalar product is a Casoratian determinant.} 
We wish to show that Slavnov's determinant expression is Casoratian 
in the free rapidities $\{x\}$ of the general state. Expanding 
$\det \Omega$, using the Cauchy-Binet identity, we obtain

\begin{eqnarray}
\label{product-form}
\det \Omega
&=&
\det
\ll
\sum_{k=1}^{N+L-1} x_{i}^{k-1} \, \kappa_{kj}
\rr
\nonumber
\\
&=&
\sum_{1 \leq k_1 < \cdots < k_N \leq N+L-1}
\det
\ll
x_{i}^{k_j-1}
\rr
\det 
\ll 
\kappa_{k_i, j} 
\rr
\nonumber
\\
&=&
\sum_{0 \leq \lambda_N \leq \cdots \leq \lambda_1 \leq L-1}
\det
\ll
x_i^{\lambda_{(N+1-j)}+j-1} 
\rr
\det
\ll
\kappa_{\lambda_{(N-i+1)}+i, j}
\rr
\nonumber
\\
&=&
\sum_{0 \leq \lambda_N \leq \cdots \leq \lambda_1 \leq L-1}
\det
\ll
x_i^{\lambda_{j}+N+1-j-1}
\rr
\det
\ll
\kappa_{\lambda_{i}+N+1-i, j}
\rr
\nonumber
\\
\end{eqnarray}

\noindent From the definition of Schur polynomials that uses 
the Jacobi-Trudi identity in Equation (\ref{Schur}) we obtain

\begin{eqnarray*}
\det \Omega' 
&=&
\frac{ 
\det \Omega
}{
\Delta\{x\}
}
=
\sum_{0 \leq \lambda_N \leq \cdots \leq \lambda_1 \leq L-1}
\det
\ll
h_{\lambda_{j}-j+i}\{x\}
\rr
\det
\ll
\kappa_{\lambda_{i}+N+1-i, j}
\rr
\\
&=&
\sum_{0 \leq \lambda_N \leq \cdots \leq \lambda_1 \leq L-1}
\det
\ll
h_{\lambda_{N+1-j}-N-1+j+i}\{x\}
\rr
\det
\ll
\kappa_{\lambda_{i}+N+1-i, j}
\rr
\\
&=&
\sum_{1 \leq k_1 \le \cdots \le k_N \leq N+L-1}
\det
\ll
h_{k_j -N-1+i}\{x\}
\rr
\det
\ll
\kappa_{k_i, j}
\rr
\end{eqnarray*}
\begin{equation}
=
\det 
\ll
\sum_{k=1}^{N+L-1} 
h_{k-N-1+i}\{x\}  
\ \ 
\kappa_{kj}
\rr
\end{equation}

\noindent Hence $\det \Omega'$ is Casoratian in $\{x\}$. Next, we 
need to show that a Casoratian determinant is a solution of the 
bilinear identities of discrete KP, but this requires a number of 
definitions which we outline in the next section.

\section{Continuous KP, Miwa variables and discrete KP.}

\noindent A standard introduction to the continuous KP hierarchy 
is \cite{blue-book}.  Miwa variables are discussed in detail in 
\cite{konopelchenko} where 
further references to their applications are provided. 
The discrete KP hierarchy was introduced in \cite{hirota}, and 
further studied in \cite{miwa-discrete} and \cite{djm2}. 
In this note, we follow the treatment in \cite{ohta}.

\paragraph{Continuous KP} is an infinite hierarchy of integrable 
partial differential equations generated in Hirota's bilinear 
form by expanding the bilinear identity

\begin{equation}
\oint_{k=k_{\infty}} \frac{d k}{2 \pi i} 
\ 
e^{\xi(t - t', k)} \tau\{t - \epsilon(k^{-1})\}
\ 
                   \tau\{t + \epsilon(k^{-1})\}
=0
\label{bilinear-differential-identity}
\end{equation}

\noindent where $k \in \mathbb{P}^1$, the contour integral is 
around the point at infinity $k_{infinity} \in \mathbb{P}^1$, 
$\{t\} = \{t_1, t_2, t_3, \cdots\}$,
$\xi(t, k) = \sum_{i=1}^{\infty} t_i k^i$,
$\epsilon(k^{-1}) = \{ \frac{1}{k}, 
                       \frac{1}{2 k^2}, 
                       \frac{1}{3 k^3}, \cdots \}$,
$\{t \pm \epsilon(k^{-1})\} = 
 \{ 
 t_1 \pm \frac{1}{ k}, 
 t_2 \pm \frac{1}{2k},
 t_3 \pm \frac{1}{3k}, \cdots \}$. The simplest KP equation in 
the hierarchy is  

\begin{equation}
\ll D_1^4 + 3 D_2^2 - 4 D_1 D_3 \rr \tau \cdot \tau = 0
\label{continuous-KP}
\end{equation}

\noindent where $D_i$ is the Hirota derivative with respect to $t_i$. 
For the precise definition of $D_i$ and that of the notation 
$\tau \cdot \tau$, see \cite{blue-book}. 

\paragraph{Continuous and discrete Miwa variables.} 
In \cite{miwa-discrete}, Miwa introduced two infinite sets of variables, 
the continuous variables $\{x\} = \{x_1, x_2, \cdots\}$, and 
the discrete   (and integer valued) variables 
$\{m\} = \{m_1, m_2, \cdots\}$, and showed that setting 

\begin{equation}
t_j = 
\sum_{i=1}^{\infty} m_i \frac{x_i^j}{j}
\label{miwa-change}
\end{equation}

\noindent transforms $\tau$-functions of continuous KP 
to $\tau$-functions of a hierarchy of bilinear difference equations,
namely discrete KP, studied in detail in \cite{djm2}. 
These variables are now known as continuous and discrete Miwa 
variables, respectively, 

\paragraph{Multiplicities.} From Equation (\ref{miwa-change}), one
can see that 
the discrete variables $\{m\}$, where $m_i \in \mathbb{Z}$ are 
multiplicities of the continuous
variables $\{x\}$. In other words, $m_i > 1$ is equivalent to saying 
that $x_i$ occurs $m_i$ times in $\{x\}$, or that there are $m_i$ continuous
variables that have the same value $x_i$.

\paragraph{Discrete KP} is an infinite hierarchy of integrable 
partial {\it difference} 
equations in an infinite set of continuous Miwa variables $\{x\}$,
where time evolution is obtained by changing the multiplicities 
$\{m\}$ of these variables. In this note, we are interested in 
situation where the total number of continuous Miwa variables is 
finite, and the sum of all multiplicities is $N$. In this case, the 
discrete KP hierarchy can be written in bilinear form as $n$$\times$$n$ 
determinant equations

\begin{equation}
\det
\ll
\begin{array}{cccccc}
     1 & x_1    & \cdots & x_1^{n-2} & \, & x_1^{n-2} \tau_{+1}\{x\}
\tau_{-1}\{x\} \\
     1 & x_2    & \cdots & x_2^{n-2} & \, & x_2^{n-2} \tau_{+2}\{x\}
\tau_{-2}\{x\} \\
\vdots & \vdots & \vdots & \vdots    & \, & \vdots                         
        \\
     1 & x_n    & \cdots & x_n^{n-2} & \, & x_n^{n-2} \tau_{+n}\{x\}
\tau_{-n}\{x\} 
\end{array}
\rr
=0
\label{bilinear-difference-equation}
\end{equation}

\noindent where $3 \leq n \leq N$, and

\begin{eqnarray}
\tau_{+i}\{x\} &=& \tau\{ x_1^{(m_1    )}, \cdots, x_i^{(m_i + 1)}, \cdots, x_N^{(m_N    )}\},
\nonumber
\\
\tau_{-i}\{x\} &=& \tau\{ x_1^{(m_1 + 1)}, \cdots, x_i^{(m_i    )}, \cdots, x_N^{(m_N + 1)}\}
\end{eqnarray}

\noindent In other words, if $\tau\{x\}$ has $m_i$ copies of the variable $x_i$, then 
$\tau_{+i}\{x\}$ has $m_i + 1$ copies of $x_i$ and the multiplicities of all other 
variables remain the same, while $\tau_{-i}\{x\}$ has one more copy of each variable 
except $x_i$. Equivalently, one can use the simpler notation

\begin{eqnarray}
\tau_{+i}\{x\} &=& \tau\{  m_1,      \cdots, (m_i + 1), \cdots,  m_N     \},
\nonumber
\\
\tau_{-i}\{x\} &=& \tau\{ (m_1 + 1), \cdots,  m_i     , \cdots, (m_N + 1)\}
\label{notation-for-tau}
\end{eqnarray}

\noindent The simplest discrete KP bilinear difference equation, in the notation 
of Equation (\ref{notation-for-tau}), is 

\begin{eqnarray}
&\phantom{+}&
x_i(x_j-x_k) \tau\{ m_i+1, m_j,   m_k   \} \tau\{ m_i,   m_j+1, m_k+1 \} 
\nonumber \\
&+& 
x_j(x_k-x_i) \tau\{ m_i,   m_j+1, m_k   \} \tau\{ m_i+1, m_j,   m_k+1 \}
\nonumber \\
&+& 
x_k(x_i-x_j) \tau\{ m_i,   m_j,   m_k+1 \} \tau\{ m_i+1, m_j+1, m_k   \}   
= 0
\label{miwa-hirota}
\end{eqnarray}

\noindent where 
$\{x_i, x_j, x_k\} \in \{x\}$ and 
$\{m_i, m_j, m_k\} \in \{m\}$ are
any two (corresponding) triples in the sets of continuous and
discrete (integral valued) Miwa variables. Equation (\ref{miwa-hirota}) 
is the discrete analogue of Equation (\ref{continuous-KP}).

\paragraph{Discrete time evolution in discrete KP.} Each continuous 
Miwa variable $x_i$ corresponds to a time variable in discrete KP. 
Time evolution in discrete KP, in direction $x_i$, is given by the 
discrete changes in the multiplicities $m_i$ of $x_i$. Notice that 
as a multiplicity $m_i$ changes by $\pm 1$, the rank of the matrix 
$M_{i \pm 1}$, where $\det M_{i \pm 1} = \tau_{i \pm 1}$ remains 
the same as the rank of $M$, where $\det M = \tau$. 

\section{Bethe scalar products are discrete KP $\tau$-functions.}

\noindent In this section, we adapt the general treatment of 
\cite{ohta} to the specific case of Slavnov's determinant 
expressions. We do this in detail to show explicitly that 
the general (and slightly abstract) identities and theorems 
in \cite{ohta} apply to Slavnov's expressions.

\paragraph{Re-arranging the elements of Slavnov's determinant.} 
Given the $N$$\times$$N$ matrix $\Omega'$ with elements 

\begin{equation}
\omega'_{ij}  = \sum_{k=1}^{N+L-1} h_{k-N-1+i}\{x\} \ \kappa_{kj} 
\end{equation}

\noindent let us consider the matrix $\Omega''$ with elements

\begin{equation}
\omega_{ij} = \sum_{k=1}^{N+L-1} c_{ik} \ h_{k-j}\{x\} 
\end{equation}

\noindent which is obtained from $\Omega'$ by reordering the 
rows of the latter from bottom to top, changing the rows
and the columns and setting $c_{ik} = \kappa_{ki}$. Notice that 
we use $\omega$ rather than $\omega''$ for the elements 
of $\Omega''$ to simplify the notation. Since
$\det \Omega' = (-)^{N(N-1)/2} \det \Omega$, it is sufficient 
to show that $\Omega$ satisfies the difference bilinear 
identities of discrete KP. 

\paragraph{Identities for the elements $\omega_{ij}$.} It 
follows from Equations (\ref{i1}--\ref{i2}) that the elements 
$\omega_{ij}$ of $\Omega''$ satisfy analogous identities

\begin{eqnarray}
\omega_{ij}\{x_1, \ldots, x_{m}^{(2)}, \ldots, x_{N}\}
&=&
\omega_{i j  }\{x_1, \ldots,                      x_N \} 
\nonumber
\\
&+& 
x_m \omega_{i,j+1}\{x_1, \ldots, x^{(2)}_{m}, \ldots, x_N \}
\label{a1}
\end{eqnarray}

\begin{multline}
(x_r - x_s) 
\  
      \omega_{ij} \{x_{1}, \ldots, x^{(2)}_{r}, x^{(2)}_{s}, \ldots x_N \} = 
\\
x_{r} \ \omega_{ij} \{x_1,   \ldots, x_{r}^{(2)}, \ldots, x_{N} \}
-
x_{s} \ \omega_{ij} \{x_1,   \ldots, x_{s}^{(2)}, \ldots, x_{N} \}
\label{a2}
\end{multline}

\noindent From Equation (\ref{discrete-derivative}), we see that  

\begin{equation}
\Delta_{m}\omega_{ij}\{x_1,\ldots,x_N\}=\omega_{i,j+1}\{x_1,\ldots,x_N\}
\end{equation}

\noindent which is equivalent to the statement that $\det \Omega$ 
is Casoratian.

\paragraph{Notation for column vectors with elements $\omega_{ij}$.} We need 
the column vector

\begin{equation}
\vec\omega_{j} =
\ll
\begin{array}{c} 
\omega_{1j} \{ x^{(m_{1})}_1, \ldots, x^{(m_{N})}_{N} \} \\ 
\omega_{2j} \{ x^{(m_{1})}_1, \ldots, x^{(m_{N})}_{N} \} \\ 
\vdots                               \\ 
\omega_{Nj} \{ x^{(m_{1})}_1, \ldots, x^{(m_{N})}_{N} \} 
\end{array}
\rr
\end{equation}

\noindent and write 

\begin{equation}
\vec\omega_{j}^{[k_1,\ldots,k_n]} =
\ll
\begin{array}{c} 
\omega_{1j}\{ x^{(m_{1})}_1, \ldots, x^{(m_{k_1}+1)}_{k_1},\ldots,x^{(m_{k_n}+1)}_{k_n},\ldots, x^{(m_{N})}_{N} \} \\ 
\omega_{2j}\{ x^{(m_{1})}_1, \ldots, x^{(m_{k_1}+1)}_{k_1},\ldots,x^{(m_{k_n}+1)}_{k_n},\ldots, x^{(m_{N})}_{N} \} \\ 
\vdots \\ 
\omega_{Nj}\{ x^{(m_{1})}_1, \ldots, x^{(m_{k_1}+1)}_{k_1},\ldots,x^{(m_{k_n}+1)}_{k_n},\ldots, x^{(m_{N})}_{N} \}
\end{array}
\rr
\end{equation}

\noindent for the corresponding column vector where the multiplicities of the variables $x_{k_1}, \cdots,
x_{k_n}$ are increased by 1. 

\paragraph{Notation for determinants with elements $\omega_{ij}$.} We also need the determinant 

\begin{equation}
\tau= 
\det 
\ll   \vec\omega_{1}\,\, 
      \vec\omega_{2}\,\,
      \cdots        \,\,
      \vec\omega_{N}
\rr
=
\big | \, 
      \vec\omega_{1}\,\, 
      \vec\omega_{2}\,\,
      \cdots        \,\,
      \vec\omega_{N}
\, \big |
\end{equation}

\noindent and the notation 

\begin{equation}
\tau^{[k_1, \ldots, k_n]} =
\\
\big | \, 
\vec\omega_{1}^{[k_1, \ldots, k_n]} \, \, 
\vec\omega_{2}^{[k_1, \ldots, k_n]} \, \, 
\cdots                           \, \, 
\vec\omega_{N}^{[k_1, \ldots, k_n]}
\, \big |
\label{defn}
\end{equation}

\noindent for the determinant with shifted multiplicities. Next, 
and closely following \cite{ohta}, we derive two identities 
involving Casoratian determinants with elements $\omega_{ij}$. 

\paragraph{Casoratian identity 1.} The first identity that we 
need is

\begin{equation}
x^{n-2}_{1} \ \tau^{[1]} =
\big | \,
\vec\omega_{1}   \,\,
\vec\omega_{2}   \,\,
\cdots           \,\, 
\vec\omega_{N-1} \,\,
\vec\omega_{N-n+2}^{[1]}
\, \big |
\label{A1}
\end{equation} 
 
\noindent which is derived as follows. From Equation (\ref{defn}), 
we have 

\begin{equation}
\tau^{[1]} =
\big | \,
\vec\omega_{1}^{[1]} \,\,
\vec\omega_{2}^{[1]} \,\,
\cdots            \,\,
\vec\omega_{N}^{[1]}
\, \big |
\end{equation}
 
\noindent In view of Equation (\ref{a1}), subtracting $x_1$ times column 
$j+1$ from column $j$ in this determinant for $j=1,2, \ldots, N-1$ 
allows us to write

\begin{equation}
\tau^{[1]} =
\big | \, 
\vec\omega_{1}   \,\,
 \vec\omega_{2}   \,\,
 \cdots           \,\,
 \vec\omega_{N-1} \,\,
 \vec\omega_{N}^{[1]}
\, \big  |
\end{equation} 

\noindent Multiplying column $N$ by $x_1$ and adding column $N-1$ 
to the result, we obtain 

\begin{equation}
x_{1} \ \tau^{[1]} =
\big | \,
\vec\omega_{1}   \,\,
\vec\omega_{2}   \,\,
\cdots           \,\,
\vec\omega_{N-1} \,\,
\vec\omega_{N-1}^{[1]}
\, \big |
\label{step1}
\end{equation} 

\noindent Similarly, multiplying column $N$ in Equation (\ref{step1}) 
by \(x_{1}\) and subtracting column $N-2$ yields

\begin{equation}
x^{2}_{1} \ \tau^{[1]} =
\big | \,
\vec\omega_{1}   \,\, 
\vec\omega_{2}   \,\,
\cdots           \,\,
\vec\omega_{N-1} \,\,
\vec\omega_{N-2}^{[1]}
\, \big |
\end{equation} 

\noindent Iterating this procedure by multiplying column $N$ by $x_1$ and 
subtracting column $N-j$, we obtain Equation (\ref{A1}).

\paragraph{Casoratian identity 2.} The second identity that we need is 

\begin{multline}
\prod_{1\leq r<s\leq n}(x_r-x_s)  \tau^{[1, \ldots, n]} =
\\
\big | \,
\vec\omega_1            \,\,
\ldots                  \,\,
\vec\omega_{N-n}        \,\,
\vec\omega_{N-n+1}^{[n]}   \,\,
\vec\omega_{N-n+1}^{[n-1]} \,\,
\ldots                  \,\,
\vec\omega_{N-n+1}^{[1]}
\, \big | 
\label{A2}
\end{multline}

\noindent which is derived as follows. From Equation (\ref{step1}), it 
follows that

\begin{equation}
x_{1}\ \tau^{[1,2]} =
\big | \,
\vec\omega_{1}^{[2]}   \,\,
\vec\omega_{2}^{[2]}   \,\,
\cdots
\vec\omega_{N-1}^{[2]} \,\,
\vec\omega_{N-1}^{[1, 2]}
\, \big |
\end{equation} 

\noindent which we can rewrite by subtracting $x_2$ times 
column $j+1$ from column $j$ for $j=1,2, \ldots,N-2$ as

\begin{equation}
x_{1}\ \tau^{[1,2]} =
\big | \,
\vec\omega_{1}      \,\,
\vec\omega_{2}      \,\,
\cdots              \,\,
\vec\omega_{N-2}    \,\,
\vec\omega_{N-1}^{[2]} \,\,
\vec\omega_{N-1}^{[1,2]}
\, \big |
\end{equation}

\noindent Multiplying column $N$ by $(x_1-x_2)$ and applying 
Equation (\ref{a2}), we see that

\begin{eqnarray}
\nonumber (x_1-x_2) x_{1} \ \tau^{[1,2]} 
&=&
x_{1}
\big | \,
\vec\omega_{1}          \,\,
\vec\omega_{2}          \,\,
\cdots                  \,\,
\vec\omega_{N-2}        \,\,
\vec\omega_{N-1}^{[2]}     \,\,
\vec\omega_{N-1}^{[1]}
\, \big |
\\
&-& 
x_{2}
\big | \,
\vec\omega_{1}          \,\,
\vec\omega_{2}          \,\,
\cdots                  \,\,
\vec\omega_{N-2}        \,\,
\vec\omega_{N-1}^{[2]}     \,\,
\vec\omega_{N-1}^{[2]}
\, \big |
\end{eqnarray}

\noindent Since the last two columns of the latter determinant 
are identical, we obtain

\begin{equation}
(x_1-x_2) \ \tau^{[1,2]} =
\big | \,
\vec\omega_1           \,\,
\ldots                 \,\,
\vec\omega_{N-2}       \,\,
\vec\omega_{N-1}^{[2]},   \,\,
\vec\omega_{N-1}^{[1]}
\, \big |
\end{equation}

\noindent which establishes Equation (\ref{A2}) for \(n=2\). 
Now suppose inductively that

\begin{multline}
\prod_{1 \leq r < s \leq n}(x_r-x_s) \ \tau^{[1, \ldots, n]} =
\\
\big | \,
\vec\omega_{1}          \ \ 
\ldots                  \ \ 
\vec\omega_{N-n}        \ \ 
\vec\omega_{N-n+1}^{[n]}   \ \ 
\vec\omega_{N-n+1}^{[n-1]} \ \ 
\cdots                  \ \ 
\vec\omega_{N-n+1}^{[1]}
\, \big |
\end{multline}

\noindent then analogously to Equation (\ref{step1}), we have

\begin{multline}
\prod_{i=1}^{n} x_i
\prod_{1 \leq r < s \leq n}(x_r -x_s) \ \tau^{[1, \ldots, n]} =  
\\
\prod_{i=1}^{n}
\big | \,
\vec\omega_{1}          \ \
\ldots                  \ \ 
\vec\omega_{N-n}        \ \ 
\vec\omega_{N-n+1}^{[n]}   \ \ 
\vec\omega_{N-n+1}^{[n-1]} \ \ 
\cdots                  \ \ 
\vec\omega_{N-n+1}^{[1]}
\, \big |  \ \ = 
\\ 
\big | \,
\vec\omega_{1}          \ \ 
\ldots                  \ \ 
\vec\omega_{N-n}        \ \ 
\vec\omega_{N-n}^{[n]}     \ \ 
\vec\omega_{N-n}^{[n-1]}   \ \ 
\cdots                  \ \ 
\vec\omega_{N-n}^{[1]}
\, \big | 
\end{multline}

\noindent It follows that

\begin{multline}
\prod_{i=1}^{n}
x_i
\prod_{1 \leq r < s \leq n}(x_r - x_s) 
\ 
\tau^{[1, \ldots, n, n+1]} = 
\\
\big | \,
\vec\omega_{1}^{[n+1]}       \ \ 
\ldots                       \ \  
\vec\omega_{N-n}^{[n+1]}     \ \ 
\vec\omega_{N-n}^{[n,n+1]}   \ \ 
\vec\omega_{N-n}^{[n-1,n+1]} \ \ 
\cdots                       \ \ 
\vec\omega_{N-n}^{[1,n+1]}
\, \big |                    \ \
=
\\
\big | \,
\vec\omega_{1}            \ \ 
\ldots                    \ \ 
\vec\omega_{N-n-1}        \ \ 
\vec\omega_{N-n}^{[n+1]}     \ \ 
\vec\omega_{N-n}^{[n,n+1]}   \ \ 
\vec\omega_{N-n}^{[n-1,n+1]} \ \ 
\cdots                    \ \ 
\vec\omega_{N-n}^{[1,n+1]}
\, \big |
\label{step2}
\end{multline}

\noindent Using Equation (\ref{a2}) repeatedly gives

\begin{multline}
\prod_{1 \leq i \leq n}(x_{i} - x_{n+1})  \ \ 
\times
\\
\big | \,
\vec\omega_{1}                           \ \ 
\ldots                                   \ \ 
\vec\omega_{N-n-1}                       \ \ 
\vec\omega_{N-n}^{[n+1]}                 \ \
\vec\omega_{N-n}^{[n,n+1]}               \ \ 
\vec\omega_{N-n}^{[n-1,n+1]}             \ \ 
\cdots                                   \ \ 
\vec\omega_{N-n}^{[1,n+1]}
\, \big |
\ \ 
=
\\ 
\prod_{i=1}^{n} x_i 
\ 
\big | \,
\vec\omega_{1}                           \ \ 
\ldots                                   \ \ 
\vec\omega_{N-n-1}                       \ \ 
\vec\omega_{N-n}^{[n+1]}                 \ \
\vec\omega_{N-n}^{[n]}                   \ \ 
\vec\omega_{N-n}^{[n-1]}                 \ \ 
\cdots                                   \ \ 
\vec\omega_{N-n}^{[1]}
\, \big |
\end{multline}

\noindent Combining this with Equation (\ref{step2}) shows 
that

\begin{multline}
\prod_{1\leq r<s\leq n+1}(x_r-x_s)\ \tau^{[1,\ldots,n+1]}
=
\\
\big | \,
\vec\omega_{1}        \ \ 
\ldots                \ \  
\vec\omega_{N-n-1}    \ \ 
\vec\omega_{N-n}^{[n]}   \ \ 
\vec\omega_{N-n}^{[n-1]} \ \ 
\cdots                \ \
\vec\omega_{N-n}^{[1]}
\, \big |
\end{multline}

\noindent thereby completing the proof of Equation (\ref{A2}). We are 
finally in a position to complete the proof that Slavnov's determinant 
expressions are discrete KP $\tau$-functions.

\paragraph{Bilinear identities from Laplace expansions.} Following \cite{ohta}, 
we consider the $2N$$\times$$2N$ determinant, which is identically zero,

\begin{multline}
\det
\ll
\begin{array}{cccccccc}
\vec\omega_1          & 
\cdots                &
\vec\omega_{N-1}      & 
\vec\omega_{N-n+2}^{[1]} &    
0_1                   & 
\cdots                &    
0_{N-n+1}             &
\vec\omega_{N-n+2}^{[n]} 
\cdots 
\vec\omega_{N-n+2}^{[2]} 
\\
0_1                   & 
\cdots                &    
0_{N-1}               & 
\vec\omega_{N-n+2}^{[1]} & 
\vec\omega_1          & 
\cdots                & 
\vec\omega_{N-n+1}    &
\vec\omega_{N-n+2}^{[n]} 
\cdots 
\vec\omega_{N-n+2}^{[2]} 
\end{array}
\rr
= 0
\\
\label{2N-by-2N}
\end{multline}

\noindent where we have used subscripts to label the zero elements 
with the positions of the columns that they are in for notational 
clarity. Performing a Laplace expansion of the left hand side of 
Equation (\ref{2N-by-2N}) in $N$$\times$$N$ minors along the top 
$N$$\times$$N$ block, we obtain 

\begin{multline}
\sum_{\nu = 1}^{n} 
(-)^{\nu - 1}
\big | \,
  \vec\omega_1 
  \cdots 
  \vec\omega_{N-1}  
  \vec\omega_{N-n+2}^{[\nu]} 
\, \big | 
\times \\
\big | \,
  \vec\omega_1 
  \cdots 
  \vec\omega_{N-n+1}         
  \vec\omega_{N-n+2}^{[n]} 
  \cdots 
  \vec\omega_{N-n+2}^{[\nu+1]} 
  \vec\omega_{N-n+2}^{[\nu-1]} 
  \cdots 
  \vec\omega_{N-n+2}^{[1]}
\, \big |
= 0
\label{Laplace-expansion}
\end{multline}

\noindent Using Equations (\ref{A1}--\ref{A2}), 
Equation (\ref{Laplace-expansion}) becomes

\begin{equation}
\sum_{\nu = 1}^{n} 
(-)^{\nu - 1}
x^{n-2}_{\nu}\ \tau^{[\nu]}
\prod_{\substack{1\leq r<s\leq n \\ r,s\neq\nu}}(x_r-x_s)\tau^{[1,\ldots\hat\nu\ldots,n]}
= 0
\label{57}
\end{equation}

\noindent which we recognise as the cofactor expansion of the determinant in Equation 
(\ref{bilinear-difference-equation}) using the last column. Hence we conclude 
that Slavnov's determinant expression for the XXZ Bethe scalar product is 
a $\tau$-function of discrete KP.

\section{Remarks.}

\paragraph{Shifted $\tau$-functions are not Bethe scalar products.} 
A Bethe scalar product that involves $m_i$ rapidities 
$x_i$, for $i = 1, 2, \cdots, i_{max}$, 
is a Casoratian determinant of a matrix of rank $r = \sum_{i=1}^{i_{max}} m_i$. 
Let us denote the corresponding $\tau$-function by
$\tau = \tau \{x_1^{(m_1)}, \cdots, x_i^{(m_i)}, \cdots, x_N^{(m_N)}\}$.
Now let us consider a time evolution of the latter, for example 
$\tau_{i+1} = \tau\{x_1^{(m_1)}, \cdots, x_i^{(m_i + 1)}, \cdots, x_N^{(m_N)}\}$.
Time evolution has increased the multiplicities by 1, but kept the rank
of the corresponding Casoratian determinant the same, thus we cannot interpret 
$\tau_{i+1}$ as a Bethe scalar product and it remains 
unclear to us how to interpret the discrete time evolution of a Bethe 
scalar product in the language of the XXZ spin chain.

\paragraph{Fermionization remains valid.} Continuous KP 
$\tau$-functions can be written as expectation values of charged 
free fermion operators \cite{blue-book}. This remains the case for
discrete KP $\tau$-functions and was the starting point of the results 
of \cite{miwa-discrete, djm2}. In \cite{fwz}, the fermion expectation 
value version of Slavnov's determinant expression 
was obtained based on an earlier result \cite{fwz-earlier}. It is 
straightforward to show that this result remains the same as the 
continuous KP time variables are restricted to be power sums of 
a finite and smaller number of continuous Miwa variables. 

\paragraph{Relation to the work of Krichever {\it et al.}} 
As mentioned earlier, our result is close in spirit to that of 
Krichever {\it et al.} \cite{krichever, zabrodin} and works
that followed including \cite{z1, z2}. The starting 
point of \cite{krichever} is that the Bethe eigenvalues
satisfy a bilinear identity that has the same structure 
as Hirota's bilinear difference equation 
and hence can be identified with $\tau$-functions of a discrete hierarchy. 
From this, a large number of interesting results follow,
including an identification of the fusion rules of the transfer matrices
of the quantum spin chain with Hirota's difference equations, that each 
step in the nested Bethe Ansatz approach to the spin chain is identified
with a classical B\"{a}cklund transformation, and most interestingly that 
the eigenvalues of Baxter's $Q$ operator are classical (suitably normalized) 
Baker-Akhiezer functions. 
On the other hand, our result is that it is the Bethe scalar 
product of a Bethe eigenstate rather than the corresponding Bethe eigenvalue 
that is identified with a discrete KP $\tau$-function, and we are far from 
obtaining further results that are analogous to those of \cite{krichever}. 
We hope that our identification is compatible with and complements that of 
\cite{krichever}.

\paragraph{Relation to the work of Sato and Sato.} Equation (\ref{57}) 
also follows from Theorem {\bf 3} of Sato and Sato \cite{Sato-Sato}. 
We didn't know this when we obtained our proof, and the existence of 
more than one proof can only shed more light on the result obtained. 

\section*{Acknowledgements.} 

\noindent OF wishes to thank M Wheeler and M Zuparic for collaboration 
on \cite{fwz, fwz-earlier} and for reading the manuscript, T Shiota 
and K Takasaki for discussions and the anonymous referee for bringing 
\cite{Sato-Sato} to our attention.
XXZ and KP are narrow examples of the broad subjects of quantum and 
classical integrability that Professor T Miwa has made major contributions 
to. We dedicate this note to him on his 60th birthday.
This work was supported by the Australian Research Council and 
a Pratt Foundation Scholarship.


\begin{thebibliography}{99}

\bibitem{wu-mccoy}
T~T~Wu, B~M~McCoy C~A~Tracy and E~Barouch, Phys Rev {\bf B13} 
(1976) 316--374

\bibitem{perk}
J~H~H~Perk,
Phys~Lett {\bf A79} (1980) 3--5.

\bibitem{au-yang-perk}
H~Au-Yang and J~H~H~Perk,
Proc of Symposia in Pure Mathematics, 
{\it Theta Functions, Bowdoin 1987}, 
L~Ehrenpreis and R~C~Gunning, Editors,
Part I, 287--294,
American Math Society, 
and references therein.

\bibitem{korepin}
A R Its, A G Izergin, V E Korepin, and N A Slavnov,
Phys Rev Lett {\bf 70} (1993) 1704--1706;
{\it Erratum} 2357--2357 

\bibitem{korepin-book}
V~E~Korepin, N~M~Bogoliubov and A~G~Izergin,
{\it Quantum inverse scattering method and correlation functions},
Cambridge University Press, (1993)

\bibitem{miwa-review}
T~Miwa,
Lecture Notes in Physics {\bf 242} (1985) 96--141,
Springer-Verlag

\bibitem{krichever}
I~Krichever, O~Lipan, P~Wiegmann and A~Zabrodin,
Comm~Math~Phys {\bf 188} (1997) 267--304

\bibitem{zabrodin}
A~V~Zabrodin, 
{\it Bethe Ansatz and classical Hirota equations},
{\tt hep-th/9607162}

\bibitem{hirota}
R~Hirota, 
J Phys Soc Jpn {\bf 50} (1981) 3785--3791  

\bibitem{kuniba}
A~Kuniba, R~Sakamoto and Y~Yamada,
Nucl~Phys {\bf B786} (2007) 207--266,
and references therein.

\bibitem{djkm}
A~Date, M~Jimbo, M~Kashiwara and T~Miwa,
in {\it Proceedings of RIMS Symposium on Non-Linear Integrable
Systems-Classical Theory and Quantum Theory}, 
M~Jimbo and T~Miwa, Editors, 
World Scientific, 1983

\bibitem{kawamoto}
N~Kawamoto, Y~Namikawa, A~Tsuchiya and Y~Yamada,
Commun Math Phys {\bf 116} (1988) 247--308

\bibitem{fwz}
O~Foda, M~Wheeler and M~Zuparic,
Nucl. Phys. {\bf B820} [FS] (2009) 649--663

\bibitem{djm2}
E~Date, M~Jimbo and T~Miwa,
J Phys Soc of Japan {\bf 51} (1982) 4125--4131

\bibitem{ohta} 
Y~Ohta, R~Hirota, S~Tsujimoto and T~Inami, 
J~of the Phys Soc of Japan, vol {\bf 62} (1993) 1872--1886

\bibitem{macdonald-book}
I~G~Macdonald,
{\it Symmetric Functions and Hall polynomials},
Oxford University Press, 2nd Edition, 1995

\bibitem{baxter-book}
R~J~Baxter,
{\it Exactly Solved Models in Statistical Mechanics},
2nd Edition,
Dover Publications, 2008

\bibitem{jimbo-miwa-book}
M~Jimbo and T~Miwa,
{\it Algebraic Analysis of Solvable Lattice Models},
CBMS Regional Conference Studies in Mathematics, 
American Math Society, 1994

\bibitem{kitanine}
N Kitanine, K K Kozlowski, J M Maillet, N A Slavnov and V Terras,
J Stat Mech (2009) P04003, and references therein.

\bibitem{slavnov}
N~A~Slavnov,
Theor Math Phys {\bf 79} (1989) 502--508

\bibitem{blue-book}
T~Miwa, M~Jimbo and E~Date, 
{\it Solitons},
Cambridge University Press, 2000

\bibitem{konopelchenko}
B~Konopelchenko and L~M~Alonso, 
Phys Lett {\bf A 258} (1999) 272--278

\bibitem{miwa-discrete}
T~Miwa,
Proc Japan Acad. {\bf 58A} (1982) 9--12

\bibitem{z1}
A~Zabrodin,
Int~J~of Mod Phys {\bf B 11} (1997) 3125--3158 

\bibitem{z2}
O~Lipan, P~Wiegmann and A~Zabrodin,
Mod~Phys~Lett {\bf A12} (1997) 1369-1378,
and 
Comm~Math~Phys {\bf 193} (1998) 373-396 

\bibitem{fwz-earlier}
O~Foda, M~Wheeler and M~Zuparic,
J~Stat~Mech (2009) P03017 

\bibitem{Sato-Sato}
M~Sato and Y~Sato,
in {\it Nonlinear Partial Differential Equations in Applied Science},
Lecture Notes in Numerical and Applied Analysis {\bf 5} (1982) 259--271.

\end{thebibliography}
\end{document}